\documentclass[12pt]{article}
\usepackage{graphicx}
\usepackage{amsfonts}
\usepackage{amssymb}
\newcommand{\nin}{\noindent}
\newcommand{\be}{\begin{equation}}
\newcommand{\ee}{\end{equation}}
\newcommand{\bea}{\begin{eqnarray}}
\newcommand{\eea}{\end{eqnarray}}

\newcommand{\nn}{\nonumber\\}

\begin{document}

\begin{center}

{\Large{\bf Liouville-Lifshitz theory in 3+1 dimensions}}
\vspace{0.5cm}

J. Alexandre$^a$\footnote{jean.alexandre@kcl.ac.uk}, K. Farakos$^b$\footnote{kfarakos@central.ntua.gr},
A. Tsapalis $^{b,c}$\footnote{a.tsapalis@iasa.gr}

\vspace{0.5cm}

$^a$ King's College London, Department of Physics, WC2R 2LS, UK

$^b$ Department of Physics, National Technical University of
       Athens \\ Zografou Campus, 157 80 Athens, Greece

$^c$ Hellenic Naval Academy, Hatzikyriakou Avenue, Pireaus 185 39, Greece

\vspace{2cm}

{\bf Abstract}

\end{center}

\nin 
We consider a four-dimensional theory in the $z=3$ Lifshitz context, with an exponential (Liouville) potential.
We determine the exact renormalized potential of the theory and derive the non-perturbative relation between the renormalized and bare couplings.
In addition, we show that Lorentz symmetry is naturally generated by quantum fluctuations in the infrared regime,
and conclude that the model can be relevant to High Energy Physics.

\vspace{2cm}

\section{Introduction}

Recently, quantum field theories in the Lifshitz context have attracted attention, exhibiting interesting renormalization properties \cite{field}.
Lifshitz type models are based on an anisotropy between space and time directions, which is characterized by the
dynamical critical exponent z, determining the properties of space-time coordinates under scale transformations: $t \to b^z t$ and 
${\bf x} \to b~ {\bf x}$.  
For $z > 1$ the higher powers of momentum in the propagators lower the superficial degree of divergence of graphs, yielding the renormalizability of 
new interactions, such as the four-fermion interaction \cite{4Fermi}. Also, divergences of renormalizable interactions in the Standard Model become softer \cite{SM}
as for example in the Yukawa model \cite{Yukawa}, where only logarithmic divergences appear.

While absent from the classical action, Lorentz symmetry is naturally generated in  Lifshitz-type models through quantum corrections, since the corresponding
kinetic term is a relevant operator and dominates the dispersion relation of the modes in the infrared regime (IR). 
Note however that recovering the speed of light, in theories with more than one species of interacting fields, requires fine-tuning of bare parameters~\cite{iengo}. 

In the context of Lifshitz-type models, the scalar field has dimensionality $[\phi]=(D-z)/2$, such that
for $D=z$ it is dimensionless. As a consequence, any power $\phi^n$ represents a classically marginal operator.
This is also the case of Liouville theory, in 1+1 dimensions, where the potential is $\mu^2\exp(g\phi)$.
It is known that, after quantization, the potential maintains its exponential form, with
renormalized parameters $\mu_r$ and $g_r$. In this theory, the renormalized coupling $g_r$ receives finite corrections and its exact relation 
to the bare coupling $g$ is known \cite{jackiw,thorn}. We show in this paper that these results hold for the $3+1$ dimensional Liouville
potential in the $z=3$ Lifshitz theory. Our proof is based on both exact functional properties and the complete resummation of diverging graphs. 
Note that this approach offers an independent derivation for the known results of the 1+1 dimensional Liouville theory.

The outline of the paper is the following. In section 2 we present the classical model and its quantization via path integral.
The exact functional form of the renormalized potential is derived in section 3. As an illustration, we calculate  the one-loop
renormalized mass in section 4, where we also show that the Lorentz-restoring kinetic term is generated in the quantum theory.

\section{Liouville - Lifshitz model}

\subsection{Classical action}

The Liouville-Lifshitz model for $D=z=3$ is defined by the classical action,
\be\label{model}
S_\mu=\int dtd^3{\bf x}
\left( \frac{1}{2}(\dot\phi)^2-\frac{1}{2}\partial^k\phi\Delta^2\partial_k\phi-\frac{\mu^6}{g^2} e^{g\phi}\right) ,
\ee
where $[\phi]=[g]=0$ and $[\mu]=1$. From naive power counting, the theory is expected to be renormalizable.
An essential property of the model (\ref{model}) is the following:
a constant shift in the field $\phi(x)\to\phi(x)+\eta$ is equivalent to a redefinition of the
only dimensionful parameter as
\be
\label{mutilde}
\mu^6\to\tilde\mu^6=\mu^6e^{g\eta},
\ee
and we will show how this property enables us to determine exactly the functional form of the 
renormalized potential of the theory.
Due to the higher order spatial derivatives in the classical action, the propagator to be used in the 
diagrammatic analysis is
\be
G(\omega, {\bf p}) = \frac{i}{\omega^2-({\bf p}^2)^3-\mu^6 +i\varepsilon} ~,
\ee
as determined by the quadratic part of the action~(\ref{model}).

\subsection{Path integral quantization}

The path integral for the model (\ref{model}) is
\be\label{path}
Z_\mu[j]=\int{\cal D}[\phi]\exp\left(iS_\mu[\phi]+i\int j\phi \right) = \exp (i W_\mu [j])
\ee
where $j$ is the source and $W_\mu[j]$ is the connected graphs generating functional. The classical field is then defined as
\be
\phi_{cl}(x)=\frac{\delta W_\mu [j]}{\delta j (x)}.
\ee
Shifting the field by a constant  $\phi(x)\to\phi(x) + \eta$ in the path integral (\ref{path}) leads to
\be\label{ZtildeZ}
W_\mu[j]=W_{\tilde\mu}[j] + \int j(x) \eta
\ee
from which a functional derivative with respect to $j(x)$ gives
\be\label{phitildephi}
\tilde\phi_{cl}(x)=\phi_{cl}(x)-\eta,
\ee
where $\tilde\phi_{cl}(x)$ is the classical field calculated with the parameter $\tilde\mu$ defined by eq.(\ref{mutilde}).
If we are interested in the effective potential only,
it is enough to consider a constant source $j_0$ from which the corresponding constant classical fields
are denoted $\phi_0$ and $\tilde\phi_0$ for
the parameters $\mu$ and $\tilde\mu$ respectively, such that $\tilde\phi_0=\phi_0-\eta$.
The partition function, regularized by the cut off $\Lambda$, is then
\be\label{ZU}
Z_\mu[j_0]=\exp\left( i\int U(\phi_0)+i\int j_0\phi_0\right),
\ee
where $U(\phi_0)$, depending on $\Lambda$, is the effective potential, defined as the derivative independent part of
the proper graphs generating functional. The latter
is the Legendre transform of $W[j_0]$, where $j_0$ has to be understood as a function
of the classical field $\phi_0$. The cut off dependence will be taken into account through the dimensionless parameter
$t=\ln(2^{1/3}\Lambda/\mu)$, where for convenience the factor $2^{1/3}$
has been absorbed in the logarithm (see the origin of this factor in eq.~(\ref{C1PL})).
It is then easy to see that the properties (\ref{ZtildeZ}) and (\ref{phitildephi}) lead to the exact identity
\be\label{UtildeU}
U(\tilde\mu,\tilde t,\tilde\phi_0)=U(\mu,t,\phi_0),
\ee
where
\be
\tilde t=\ln\left(\frac{2^{1/3}\Lambda}{\tilde\mu}\right)=t-\frac{g\eta}{6},
\ee
and the dependence on $g$ is understood on both sides of eq.(\ref{UtildeU}).

As a consequence,
the effective potential $U$ must be a function
of invariant combinations of $\mu,t,\phi_0$ as these parameters change to $\tilde\mu,\tilde t,\tilde\phi_0$, 
and the only possibility is
\be\label{UF}
U=\frac{\mu^6}{g^2}e^{g \phi_0}F(z),
\ee
where $F$ is a function of $z=g\phi_0-6t$  ($F$ can also depend on $g$, independently of $z$).
Also, the invariance expressed in eq.(\ref{UtildeU}) leads to 
\bea
\label{dUdeta}
0&=&\left( \frac{d U(\tilde\mu,\tilde t,\tilde\phi_0)}{d\eta}\right)_{\eta=0}= 
\left( \frac{\partial U}{\partial\tilde\mu^6} \frac{\partial\tilde\mu^6}{\partial \eta} + 
 \frac{\partial U}{\partial\tilde t} \frac{\partial\tilde t}{\partial \eta} +
  \frac{\partial U}{\partial\tilde \phi_0} \frac{\partial\tilde \phi_0}{\partial \eta} \right)_{\eta=0} \nn
&=&g\mu^6\frac{\partial U}{\partial\mu^6}-\frac{g}{6}\dot U-U',
\eea
where a prime denotes a derivative with respect to $\phi_0$ and a dot denotes a derivative with respect to $t$.
Together with eq.(\ref{UF}) we obtain
\be\label{Uprime}
U'=gU-\frac{g}{6}\dot U,
\ee
As will be seen in the
next section, the partial differential equation (\ref{Uprime}) will lead us to the exact field dependence of the effective potential $U$, 
in the limit where $t\to\infty$.

\subsection{Loop expansion}

Quantum corrections to the potential are calculable in the loop expansion by standard methods~\cite{JackiwLoop} by summing
all the vacuum diagrams of the theory, leading to an expansion of $F(z)$ in $\hbar$. We show here that the loop expansion is
simultaneously an expansion in $g^2$.\\
The perturbative treatment of the model requires the expansion of the potential in power
series of the coupling $g$ and therefore generates an infinite series of vertices. Besides the tadpole $\mu^6\phi/g$
and the mass term $\mu^2\phi^2/2$,  n-point vertices $\gamma_n$ are generated by the expansion of the exponential
\be
\gamma_n =\mu^6 g^{n-2} ~~~~~~~~n \ge 3.
\label{vn}
\ee
The number $L$ of loops of a given vacuum graph is related to the number $P$ of propagators and the number $V$ of vertices
by
\be
L = P - V + 1 ~.
\ee
If $V_n$ denotes the number of vertices with $n \ge 3$ legs, we have
\be
V = \sum_{n=3}^N  V_n , ~~~~{\rm and}~~~~ \sum_{n=3}^N n V_n = 2 P~,
\ee
where $N$ is the highest number of legs joining at the same vertex, in the specific graph which is considered.
Since a vertex with $n$ legs is proportional to $g^{n-2}$, the vacuum graph is proportional to
a power of $g$ equal to
\be
\sum_{n=3}^N (n-2)V_n = 2 P - 2 V = 2 L - 2 ~.
\ee
Hence, if we consider the factor $g^{-2}$ in eq.(\ref{UF}), one can see that a $L$-loop graph in the expansion of the
potential $U$ is proportional to $(g^2)^L$: the expansion in $\hbar$ is equivalent to an expansion in $g^2$.

\section{Exact renormalized potential}

\subsection{Diagrammatic analysis and counterterms}

By power counting, one can see that the {\it only} source of divergence in this model
is a loop made of one propagator only, that
we denote 1PL for ``one propagator loop'', and which is equal to
\bea\label{C1PL}
C_{1PL} &=& \int_{-\infty}^{+\infty} \frac{d\omega}{2\pi}
\int \frac{d^3 p}{(2\pi)^3}
\;\frac{i}{\omega^2 -p^6-\mu^6} = \frac{1}{2} \int \frac{d^3 p}{(2\pi)^3}
\frac{1}{\sqrt{p^6 + \mu^6}} \nn
&=& \frac{1}{12\pi^2}~{\rm sinh}^{-1}\left(\frac{\Lambda^3}{\mu^3}\right)
\simeq \frac{1}{12\pi^2}\ln\left( 2\frac{\Lambda^3}{\mu^3}\right)
\equiv\frac{t}{4\pi^2}
\eea
The ``1PL''  loop may appear a multitude of times adjacent to a
n-point vertex (see Fig.1), in fact up to k times as long as $2k \le n$.
A n-point vertex will be responsible therefore for divergences coming from
the appearance of k ``1PL'' loops, for $k \le [\frac{n}{2}]$. 

\vspace{1cm}

\begin{figure}[ht]
\begin{center}
\includegraphics[width=12cm]{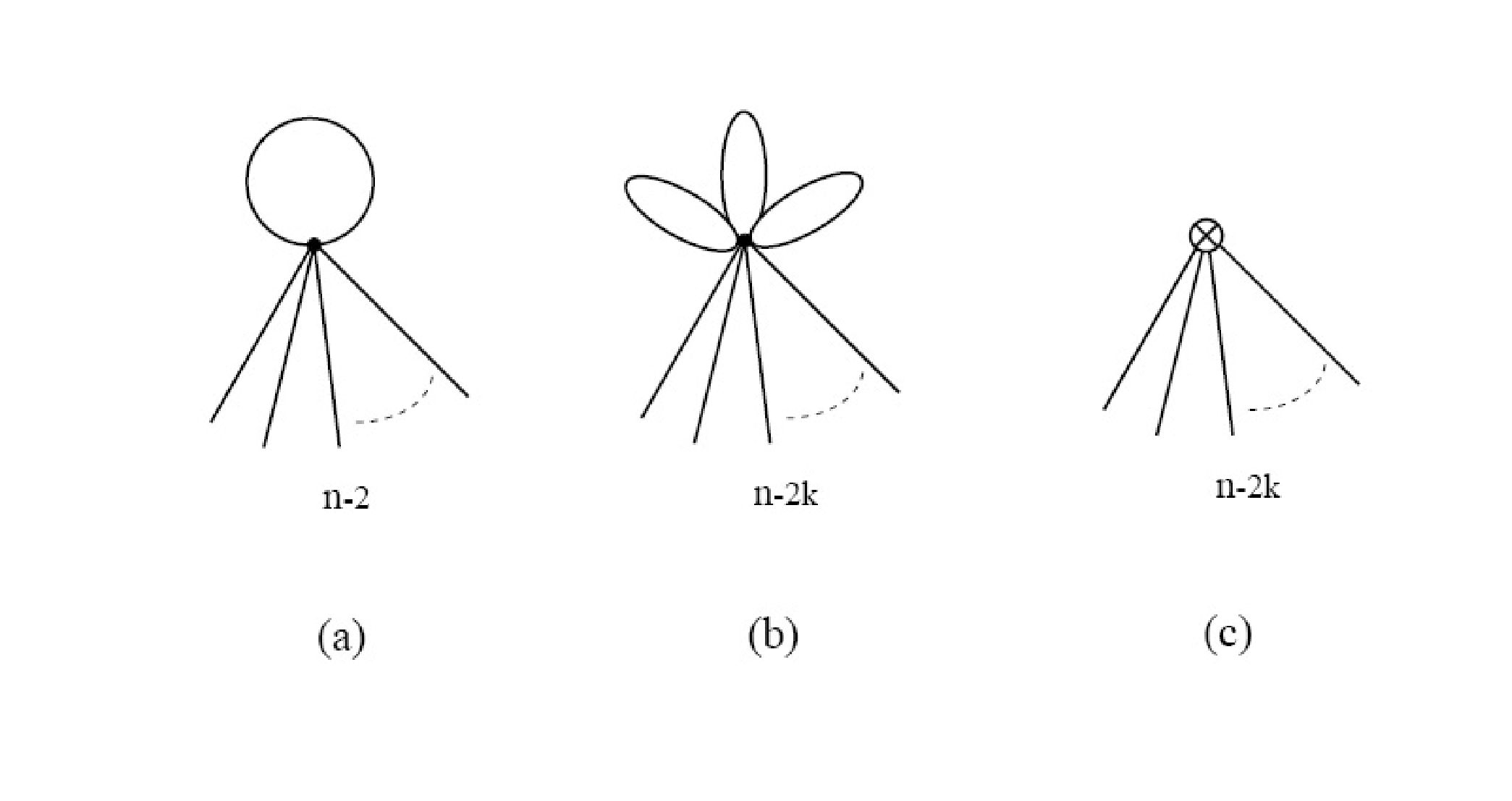}
\caption{Diagrams contributing to infinities in the Liouville theory
from the n-point vertex: (a) one loop infinity renormalized by an (n-2)-point
counterterm. (b) k-loop infinity renormalized by the (n-2k)-point countertem
(c)}
\end{center}
\end{figure}

\nin We now derive the structure of the countertems of the theory, using a proof by induction.\\
\nin{\it (i) One-loop.}\\
Counting the number of ways to stick one 1PL to a $n$-point vertex, the one loop divergence corresponding to a $n$-point function is
\be
C_n^{(1)} = \frac{\mu^6 g^{n-2}}{n!} ~\frac{n(n-1)}{2}~C_{1PL},
\ee
and the summation over the vertices in the corresponding effective potential gives
\be
\sum_{n=2}^{\infty} ~ C_n^{(1)}\phi_0^{n-2} = \sum_{n=2}^\infty\frac{\mu^6 g^{n-2}}{(n-2)!}\frac{C_{1PL}}{2}\phi_0^{n-2} =
\frac{\mu^6 t}{8\pi^2}~{e}^{g\phi_0}
\ee
In order to cure this divergence, in the minimal substraction scheme, we add the one-loop counterterm to the bare mass parameter $\mu^6\to\mu^6+\delta\mu^6_{(1)}$, with
\be\label{mu61}
\delta \mu^6_{(1)} = -\mu^6 \frac{g^2t}{8\pi^2}
\ee
{\it (ii) k-loops.}\\
A $k$-loop diagram contains at most $k$ 1PLs, and those containing less than $k$ 1PLs are cured by the $(k-1)$-loop counterterms.
As a consequence, the new divergence appearing at $k$ loops is carried by graphs where all the 1PLs are stuck at {\it the same
vertex}. Such a $n$-point vertex (see fig.1b) contributes with a divergence
\be
C_n^{(k)} = \frac{\mu^6 g^{n-2}}{n!} \,\frac{n(n-1)(n-2)\dots(n-2k+1)}
{2^k \, k!} \,(C_{1PL})^k ~~.
\ee
The summation of the dominant divergences over the vertices $n$ contributes to the corresponding effective potential
\be
\sum_{n=2k}^{\infty} ~  C_n^{(k)}\phi_0^{n-2k} =
\sum_{n=2k}^{\infty}~ \frac{\mu^6 g^{n-2}}{(n-2k)!}~\frac{1}{k!}\left(\frac{C_{1PL}}{2}\right)^k\phi_0^{n-2k}
=\frac{\mu^6}{g^2k!}\left(\frac{g^2t}{8\pi^2}\right)^k~{e}^{g\phi_0}
\ee
In order to cure this divergence, in a minimal substraction scheme, we add the $k$-loop counterterm to the $(k-1)$-loop
mass parameter $\mu^6_{(k-1)}\to\mu^6_{(k-1)}+\delta\mu^6_{(k)}$, with
\be
\delta \mu^6_{(k)} = -\frac{\mu^6}{k!}\left(\frac{g^2t}{8\pi^2}\right)^k,
\ee
{\it (iii) Complete resummation}\\
The complete cancellation of infinities requires therefore the introduction  of the {\it unique} counterterm (up to finite terms)
\be\label{Uct}
U_{ct}=\sum_{k=1}^{\infty}~\frac{\delta \mu^6_{(k)}}{g^2}  ~{e}^{g\phi_0}
= \frac{\mu^6}{g^2}  \left[1-\exp\left(\frac{\hbar g^2t}{8\pi^2}\right) \right]~e^{g \phi_0},
\ee
where the factor $\hbar$ is explicitly written, to emphasize that the loop expansion is  also an expansion in $g^2$.
We conclude that all the divergences are contained in the mass parameter of the model, and that the coupling $g$ receives finite
quantum corrections, which we now calculate.

\subsection{Renormalized coupling and mass parameter}

We write the effective potential as the sum of the contributions $U^{(k)}$ from each loop order $k$:
\be
U=U_{bare}+\sum_{k=1}^\infty U^{(k)},
\ee
where the bare potential is $U_{bare} \equiv U^{(0)}$, and $U^{(k)}$ is a polynomial of $t$:
\be
U^{(k)}=\sum_{l=0}^{k} a_l^{(k)}(\phi_0)~t^l.
\ee
Although it is not possible to know exactly all the coefficients $a_l^{(k)}$, we know the dominant divergence $a_k^{(k)}~t^k$
from the previous discussion, and one can write for $k\ge1$
\be
U^{(k)}=\frac{\mu^6}{g^2k!}\left( \frac{g^2t}{8\pi^2}\right) ^ke^{g\phi_0}~[1+{\cal O}(t^{-1})],
\ee
where the orders ${\cal O}(t^{-1})$ also depend on $\phi_0$. As a consequence, we have for $k\ge 1$,
\be
\dot U^{(k)}=\frac{g}{8\pi^2}\frac{\partial }{\partial\phi_0}U^{(k-1)}~[1+{\cal O}(t^{-1})].
\ee
Taking into account $\dot U^{(0)}=\dot U_{bare}=0$, the summation over the loops gives
\be
\dot U=\frac{g}{8\pi^2}U'~[1+{\cal O}(t^{-1})]~.
\ee
Substituting this result in eq.(\ref{Uprime}) leads to 
\be
\left(1+\frac{g^2}{48 \pi^2} \right)U' = g\,U~[1+{\cal O}(t^{-1})]~,
\ee
which integrates as
\be\label{Ufinal}
U=C(t)~ \exp(g_r\phi_0)~[1+{\cal O} (t^{-1})]~,
\ee
where $C(t)$ does not depend on $\phi_0$ and
\be\label{gr}
g_r\equiv g\left(1+\frac{g^2}{48\pi^2}\right)^{-1}~.
\ee
This relation between $g_r$ and $g$ is exact, and gives the field dependence of the renormalized potential.
In order to completely determine the latter, one still needs to specify the renormalized mass parameter $\mu_r^6$.
This is done by following the usual procedure, which consists in adding the counterterm (\ref{Uct}) to the bare potential:
\be
U_{bare}+U_{ct}=\frac{\mu^6}{g^2}\left( 2-\exp\left( \frac{\hbar g^2t}{8\pi^2}\right)\right)  ~e^{g\phi_0},
\ee
and, loop after loop, removing divergences perturbatively in order to obtain the (finite) renormalized potential $U_r=A \exp(g_r\phi_0)$ 
in the limit $t \to \infty$.
The constant $A$ can be determined in perturbation theory only, since it contains all the finite graphs of the theory.
Identifying $A$ with $\mu_r^6/g_r^2$ defines the renormalized mass parameter $\mu_r^6$, such that the renormalized potential is finally
\be\label{Ur}
U_r=\frac{\mu_r^6}{g_r^2}\exp(g_r\phi)~.
\ee
Note that, in general, $\mu_r^6$ depends on the substraction scheme, and is fixed here by the minimal substraction scheme in which the 
counterterm is given by eq.(\ref{Uct}). In the next section, we will explicitly calculate the one-loop renormalized mass parameter in this scheme.

\section{One-loop theory}

In this section, we first illustrate the result (\ref{Ur}) with the explicit calculation of the one-loop effective potential, which
determines the one-loop renormalized mass parameter $\mu_r^6$. 
We then calculate the one-loop kinetic term quadratic in derivatives, showing the restoration of Lorentz symmetry in the IR of the 
quantum theory.

\subsection{One-loop effective potential}

The one-loop effective potential is
\be
U^{(1)}=\frac{\mu^6}{g^2}e^{g\phi_0}+\frac{1}{2}\int\frac{d\omega}{2\pi}\frac{d^3{\bf p}}{(2\pi)^3}
\ln\left( \frac{\omega^2+({\bf p}^2)^3+\mu^6e^{g\phi_0}}{\omega^2+({\bf p}^2)^3+\mu^6}\right),
\ee
and it is easy to calculate its field derivative
\bea
\frac{\partial U^{(1)}}{\partial\phi_0}&=&\frac{\mu^6}{g}e^{g\phi_0}+\frac{\mu^6ge^{g\phi_0}}{8\pi^3}
\int p^2dp\int\frac{d\omega}{\omega^2+p^6+\mu^6e^{g\phi_0}}\nn
&=&\frac{\mu^6}{g}e^{g\phi_0}+\frac{\mu^6ge^{g\phi_0}}{24\pi^2}\sinh^{-1}\left( \frac{\Lambda^3}{\mu^3e^{g\phi_0/2}}\right) \nn
&=&\frac{\mu^6}{g}e^{g\phi_0}+\frac{\mu^6ge^{g\phi_0}}{8\pi^2}\left( t-\frac{g\phi_0}{6}\right)
+{\cal O}(\mu/\Lambda)^2,
\eea
such that, ignoring terms vanishing in the limit $\Lambda\to \infty$,
\be\label{U1}
U^{(1)}=\frac{\mu^6}{g^2}e^{g\phi_0}\left\lbrace 1+\frac{g^2}{8\pi^2}\left( t-\frac{g\phi_0}{6}+\frac{1}{6}
\right)\right\rbrace.
\ee
The field dependence in the expression (\ref{U1}) can be written
\be
e^{g\phi_0}\left( 1-\frac{g^3}{48\pi^2}\phi_0\right) =\exp\left\lbrace \left( g-\frac{g^3}{48\pi^2}\right) \phi_0\right\rbrace +{\cal O}(g^4),
\ee
which corresponds to the one-loop approximation of the renormalized coupling in eq.(\ref{gr}), since
\be\label{gr1}
g_r=g\left( 1+\frac{g^2}{48\pi^2}\right)^{-1}=g-\frac{g^3}{48\pi^2}+{\cal O}(g^5),
\ee
As far as the one-loop renormalized mass is concerned, from eq.(\ref{U1}) one can see that the addition of the one-loop counterterm
$U_{ct}^{(1)}=g^{-2}\delta\mu^6_{(1)}\exp(g\phi_0)$, where $\delta\mu^6_{(1)}$ is given in eq.(\ref{mu61}), leads to the definition
\be
\frac{\mu_r^6}{g_r^2}\equiv\frac{\mu^6}{g^2}\left( 1+\frac{g^2}{48\pi^2}\right)+{\cal O}(g^2),
\ee
such that
\be
\mu_r^6=\mu^6\left( 1-\frac{g^2}{48\pi^2}\right) +{\cal O}(g^4),
\ee
where the relation (\ref{gr1}) was used.

\subsection{Lorentz symmetry restoration}

Finally, we calculate the one-loop kinetic term $\partial^k\phi\partial_k\phi$
generated by quantum fluctuations, such that the IR effective theory exhibits Lorentz symmetry:
\be\label{Lorentz}
S_{eff}=\int dtd^3x\left( \frac{1+\zeta_0}{2}(\dot\phi)^2-\frac{\zeta_1}{2}\mu^4\partial^k\phi\partial_k\phi-\frac{\mu_r^6}{g_r^2}e^{g_r\phi}
+{\cal O}(\partial^4)\right),
\ee
where $\zeta_0={\cal O}(\hbar)$ and $\zeta_1={\cal O}(\hbar)$ are generated dynamically.\\
The one-loop Feynman graph responsible for the generation of this kinetic term arises from the insertion of two three-point vertices
in the propagator (the interaction $\mu^2g\phi^3$ in the Liouville potential).
Indeed, the tadpole $\mu^6\phi/g$ and the interaction $\mu^6g^2\phi^4$ give corrections
independent of the external momentum ${\bf k}$, and renormalize the mass only.
Also, the higher orders $\mu^6g^{n-2}\phi^n$ do not contribute at one-loop.
This graph is, 
\be
\frac{(-ig\mu^6)^2}{2}\int\frac{d\omega}{2\pi}\frac{d^3{\bf p}}{(2\pi)^3}\frac{i}{\omega^2-({\bf p}^2)^3-\mu^6 +i\varepsilon}~
\frac{i}{(\omega +\nu)^2-(({\bf p}+{\bf k})^2)^3-\mu^6+i\varepsilon} ~.
\ee
The contribution proportional to $\nu^2$ is (after a Wick rotation)
\bea
&&i \zeta_0 \nu^2 \\
&=&\frac{ig^2\mu^{12}}{2(2\pi)^4}\int d\omega d^3{\bf p}\left(
\frac{-\nu^2}{(\omega^2+({\bf p}^2)^3+\mu^6)^3}
+\frac{4 \omega^2 \nu^2}{(\omega^2+({\bf
p}^2)^3+\mu^6)^4}\right) \nn
&=&-\frac{ig^2\mu^{12}}{64\pi^2}\nu^2\int
\frac{p^2dp}{(p^6+\mu^6)^{5/2}} = -i\frac{g^2}{288\pi^2}\nu^2,\nonumber 
\eea 
and the contribution proportional to $k^2$ is 
\bea
&&- i \zeta_1\mu^4 k^2 \\
&=&\frac{ig^2\mu^{12}}{2(2\pi)^4}\int d\omega d^3{\bf p}\left(
\frac{-7({\bf p}^2)^2{\bf k^2}}{(\omega^2+({\bf p}^2)^3+\mu^6)^3}
+\frac{36({\bf p}^2)^4 ({\bf p}\cdot{\bf k})^2}{(\omega^2+({\bf
p}^2)^3+\mu^6)^4}\right) \nn
&=&-\frac{21ig^2\mu^{12}}{64\pi^2}k^2\int
\frac{p^6dp}{(p^6+\mu^6)^{5/2}}
+\frac{15ig^2\mu^{12}}{32\pi^2}k^2\int
\frac{p^{12}dp}{(p^6+\mu^6)^{7/2}}\nn &=&-i\frac{3g^2
I}{64\pi^2}\mu^4k^2,\nonumber \eea 
where \be I=\int_0^\infty
du\frac{7 u^6- 3 u^{12}}{(1+u^6)^{7/2}} = \frac{4}{9 \sqrt{\pi}}
\Gamma(4/3) \Gamma(13/6)\simeq 0.242. \ee We therefore see that
the Lorentz symmetric effective action (\ref{Lorentz}) is indeed
generated dynamically by quantum fluctuations, with $\zeta_0 = -g^2/(288\pi^2)$ and
$\zeta_1=3g^2 I/(64\pi^2)$. 
The dispersion relation for the low momentum modes reads
\be
\tilde{\omega}^2 = m^2 + \tilde{k}^2 + {\cal O}\left( \tilde{k}^4/{\mu}^2\right)  ~ ,
\ee
where frequency and momentum are appropriately rescaled as 
\be
\tilde{\omega}= \frac{\omega}{\mu^2} \sqrt{1+\zeta_0} ~~~,~~~ 
\tilde{k}= k \sqrt{\zeta_1} ~~~~ {\rm and}~~~~ m^2 = \frac{\mu_r^6}{\mu^4}
\ee
Hence, Lorentz violating effects are suppressed in the IR by powers of $
(\tilde k/\mu)^2 $.

\section{Conclusions}

In this work, we studied a renormalizable Lifshitz-type model with the Liouville potential, in 3+1 dimensions, and 
we demonstrated that the exponential form of the potential does not change after quantization. 
An important result is the exact
relation (\ref{gr}) between the bare and renormalized couplings, which can also be written in the form
\be
\frac{1}{g_r}=\frac{1}{g}+\frac{g}{48\pi^2}~.
\ee
This result is similar to the one obtained in the 1+1 dimensional Liouville theory \cite{jackiw,thorn}, where the factor $48\pi^2$ is replaced by $8\pi$.
Furthermore, the steps in our derivation of the renormalized potential can be repeated exactly for the 1+1 dimensional Liouville theory.
The possibility to obtain the above features arises from the use of exact functional properties and the complete resummation of graphs, which is 
specific to the exponential form of the Liouville potential.

In addition to the knowledge of the exact renormalized coupling, 
the IR regime of the quantum theory exhibits Lorentz symmetry, with a relativistic dispersion relation induced dynamically. 
An open question concerns
the possibility to have exact results for the different kinetic terms of the renormalized theory, beyond one-loop. Such a study has been done
in \cite{AKM} for the 1+1 dimensional Liouville theory, 
where it is shown, in the framework of a gradient expansion valid to all loop orders,
that the wave function renormalization vanishes. A similar study in the Lifshitz context might
also lead us to interesting properties, and this is left for a future work.

An interesting possibility is to study the Liouville-Lifshitz model in curved space time, in the context of Horava-Lifshitz gravity
\cite{Horava}. This can address a variety of cosmological questions \cite{Lifshitz+scalar}, 
in particular in the framework of quintessence models, where the exponential potential 
was first studied in \cite{quintessence}. 

Finally, the lack of a translationally invariant ground state for the quantum Liouville field theory \cite{jackiw} motivates
the extention of our study to potentials which involve more than one exponentials, where a ground state can exist.

\vspace{1cm}

\nin{\bf Aknowledgements:} We thank Nikos Tracas and Paulos Pasipoularides for useful discussions. This work is partly supported by the 
Royal Society, UK.

\end{document}